\documentclass{PoS}

\title{Tracing the merger rate of the Universe with APERTIF and ASKAP}

\ShortTitle{Extragalactic OH masers}

\author{\speaker{J. P. McKean}\\
        ASTRON, Oude Hoogeveensedijk 4, 7991 PD Dwingeloo, the Netherlands\\
        E-mail: \email{mckean@astron.nl}}

\author{A. L. Roy\\
        Max-Planck-Institut f\"{u}r Radioastronomie, Auf dem H\"{u}gel 69, D-53121 Bonn, Germany\\
        E-mail: \email{aroy@mpifr-bonn.mpg.de}}

\abstract{OH maser emission at 1.67 GHz is known to be associated with regions of intense star formation within ULIRGs. As these galaxies are formed through violent mergers, studying the co-moving density of OH maser galaxies across cosmic time will allow the merger rate of the Universe to be determined in an independent way. This merger rate is an important parameter in galaxy formation and evolution scenarios. The sensitivity, wide field of view and spectral coverage of both APERTIF on the WSRT and ASKAP will allow for the first time all-sky blind surveys for OH maser galaxies to be carried out to redshift 1.4. We describe the prospects for such surveys, including the expected number of OH maser galaxies that will be discovered, and what limits can be placed on the OH maser luminosity function, and hence the merger rate out to redshift 1.4 with various survey strategies.}

\FullConference{Panoramic Radio Astronomy: Wide-field 1-2 GHz research on galaxy evolution - PRA2009\\
		 June 02 - 05 2009\\
		 Groningen, the Netherlands}

\begin{document}

\section{Introduction}

\begin{figure}[tbh]
\begin{center}
\includegraphics[width=\textwidth]{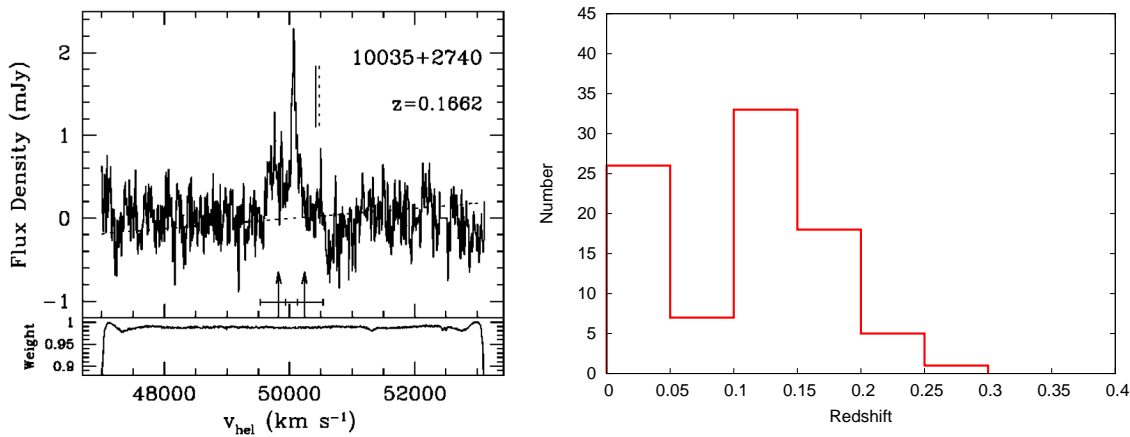} 
\caption{Left: an example of the 1.665 GHz and 1.667 GHz doublet transitions of hydroxyl from the galaxy 10035+2740 at redshift 0.1662, taken from \cite{darling02a}. Right: the redshift distribution of extragalactic OH maser galaxies. Note that this redshift distribution is not complete, but is based on targeted observations of candidate OH maser galaxies, for example, ULIRGs.} 
\label{maser-red}
\end{center} 
\end{figure}

The 1.665 GHz and 1.667 GHz doublet transitions of hydroxyl (OH) show spectacularly luminous maser emission in regions of intense circumnuclear star-formation, with reported line luminosities of up to $10^4~L_{\odot}$ (for a review see \cite{lo05}; for an example of the OH doublet maser lines see Figure \ref{maser-red}). The masing action results from a strong far-infrared radiation field, generated by the heated dust associated with star-formation, that pumps the OH molecules into an excited state \cite{baan85,baan89}. The most powerful OH masers are found in the population of ultra-luminous infrared galaxies (ULIRGs; $L_{FIR} \geq 10^{12}~L_{\odot}$) that are believed to form through galaxy mergers. Indeed, surveys of ULIRGs at low redshifts have found up to one third to show OH maser emission \cite{darling02a}, with a correlation found between the far-infrared and OH maser luminosities ($L_{OH} \propto L_{FIR}$$^{2.3}$; \cite{kloeckner04}). Thus, one of the most powerful applications of OH masers is their unbiased ability to trace the number density of ULIRGs and hence the galaxy merger history of the Universe, which forms a fundamental part of galaxy formation and evolution scenarios (e.g. \cite{croton06}). There are to date $\sim$100 OH maser galaxies known and all are found in the local Universe ($z \leq 0.3$; see Figure \ref{maser-red} for the OH maser redshift distribution).  By detecting OH masers at higher redshifts, it will be possible to determine the density evolution of the OH maser luminosity function, and hence make an independent measurement of the merger rate of the Universe as a function of redshift.  In the future, wide-field surveys with the next generation of radio telescopes and arrays can be carried out to determine the evolution in the OH maser luminosity function. The potential for using OH maser galaxies in this way was pointed out by Briggs \cite{briggs98}.

In this paper, we discuss the prospects for such surveys with two of these new facilities; the first is APERTIF (APERture Tile In Focus) on the Westerbork Synthesis Radio Telescope and the second is ASKAP (the Australian Square Kilometre Array Pathfinder).

\section{The extragalactic OH maser luminosity function}
\label{lum-func}

The OH maser luminosity function describes the co-moving number density of OH maser galaxies with luminosity $L$ per logarithmic interval in $L$. To determine the expected number of OH-emitting galaxies that will be detected by a survey, we simply integrate the OH maser luminosity function over the volume that is searched, down to the luminosity detection limit of the survey. Therefore, we clearly need an accurate form of the luminosity function. This requires a complete, or statistically well defined sample of OH maser galaxies over as large a range of maser luminosities as possible. The largest statistical sample of OH maser galaxies was obtained by Darling \& Giovanelli \cite{darling02a}, who observed $\sim150$ (U)LIRGs between redshifts $\sim 0.1$ to 0.3 with the Arecibo radio telescope, finding 50 OH maser galaxies. Using this sample, Darling \& Giovanelli \cite{darling02b} constructed the `local' OH maser luminosity function, finding a co-moving density of 
\begin{equation}
\Phi = (9.8_{-7.5}^{+31.9} \times 10^{-6})~L_{\rm OH}^{-0.64\pm0.21}~{\rm Mpc}^{-3}~{\rm dex}^{-1}.
\end{equation}
However, this luminosity function is well defined only over the luminosity range $10^{2.2}~L_{\odot} \leq L_{\rm OH} \leq 10^{3.8}~L_{\odot}$. Given that APERTIF and ASKAP will be targeting OH maser galaxies at the highest redshifts ($z<1.4$), any change in the slope of the luminosity function for $L_{\rm OH} \geq 10^{3.8}~L_{\odot}$ will clearly affect the detection rates for blind OH surveys. For our simulations, we assume the luminosity function of Darling \& Giovanelli is valid at high luminosities. However, the poorly sampled high-luminosity end of the OH maser luminosity function is currently the largest source of uncertainty in any of the predictions we make for the expected numbers of OH maser galaxies detected with APERTIF or ASKAP.

\section{Predictions for APERTIF and ASKAP}
\label{sim}

APERTIF and ASKAP are two of the new generation radio arrays that will be built during the next few years. The main science drivers of these telescopes are i) to observe H~{\sc i} beyond the local Universe, ii) to study magnetic fields in our own and nearby galaxies, and iii) carry-out deep continuum surveys of the radio sky between 0.7~GHz and 1.7~GHz. To achieve these goals, both APERTIF and ASKAP will have large instantaneous bandwidths and will be equipped with focal plane arrays to provide a much larger field of view. In Table \ref{setup} we list the design specifications for APERTIF and ASKAP that are relevant for our OH maser number-counts simulations.

To calculate the number of OH masers that can potentially be detected, we have assumed a typical maser line width of 150~km\,s$^{-1}$ \cite{darling02a}; with this line width and the line sensitivity of both telescopes, we calculated the luminosity detection limit as a function of redshift. We then integrated the luminosity function down to this luminosity detection limit and over the volume of sky defined by the field of view surveyed by the telescopes. For simplicity, we have used a total integration time of 2160 hours (3 months) for an OH survey with both APERTIF and ASKAP to allow for an easy comparison. We have also used a 12 hour integration time per pointing, but in principle this requirement can be relaxed for ASKAP since it will have much better instantaneous {\it uv}-coverage. Note that due to the 300 MHz bandwidth available, we will need to integrate three or four times per pointing to cover the full frequency and hence redshift space that is observable with both telescopes, and so each frequency setting has one third or one quarter of that total integration time. 

\begin{table}
\begin{center}
\begin{tabular}{llll}		
												&	APERTIF		& 	ASKAP	&						\\
Beam size										&	15			&	15		&	arcsec 				\\
Field of view										&	8			&	30		&	deg$^2$				\\
Bandwidth										&	300			&	300		&	MHz					\\
Frequency coverage									&	0.9--1.7		&	0.7--1.7	&	GHz					\\
Redshift range										&	0--0.86		&	0--1.39	&						\\
Line sensitivity (150 km\,s$^{-1}$ width; 12 hr; 1$\sigma$)	&	0.15			&	0.21		&	mJy\,beam$^{-1}$		\\
\end{tabular}
\end{center}
\caption{The expected specifications of APERTIF and ASKAP. These parameters were used for our OH maser number-counts simulations.}
\label{setup}
\end{table}

\begin{figure} 
\begin{center}
\includegraphics[width=\textwidth]{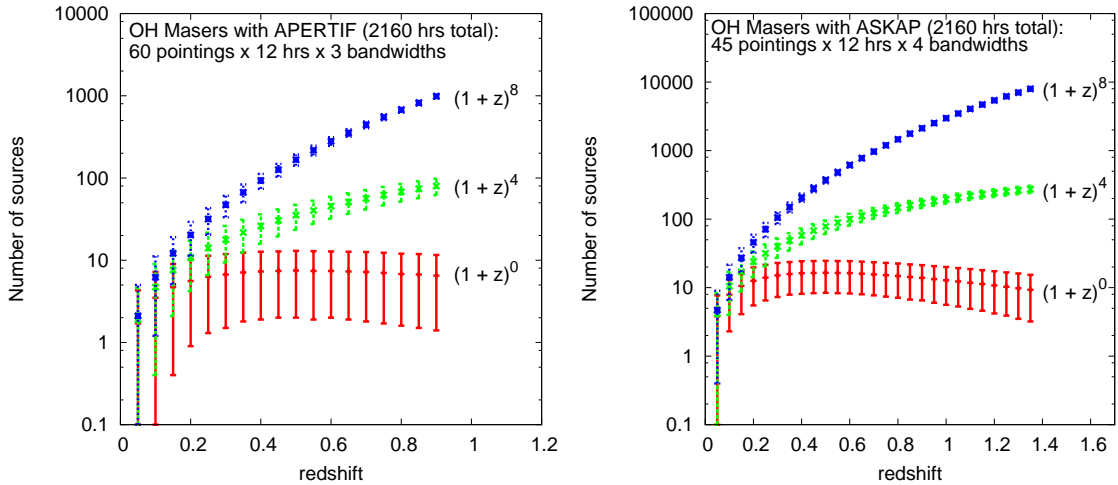} 
\caption{The predicted number of OH maser galaxies that are expected to be found with APERTIF (left) and ASKAP (right) for various evolutionary scenarios. The uncertainties are at the $2\sigma$ level.} 
\label{predictions}
\end{center} 
\end{figure}

Our results are presented in Figure \ref{predictions}, where we show the number of OH maser galaxies that are predicted for various evolutionary scenarios, $(1+z)^m$, where $m=0$ for no evolution, $m=4$ for moderate evolution (consistent with the merger-rate determined from observations of close galaxy pairs in the HUDF; \cite{ryan08}) and $m=8$ for extreme evolution. We see that one can clearly distinguish between evolutionary models above redshift 0.4 for both telescopes. Although the number of objects found per redshift bin is similar for APERTIF and ASKAP, the larger observable volume of the Universe that can be observed with ASKAP results in over an order of magnitude difference between the total number of OH masers found with both telescopes, using the same amount of observing time. Note that if the OH maser luminosity function does not follow a single power-law as assumed here, but has a knee around $10^{3}~L_{\odot}$ followed by an exponential cut-off, then the total number of predicted OH masers falls drastically to around a few tens for the most extreme evolutionary case.

\section{Survey Strategies}
\label{disc}

\begin{figure}
\begin{center}
\includegraphics[width=7.7cm]{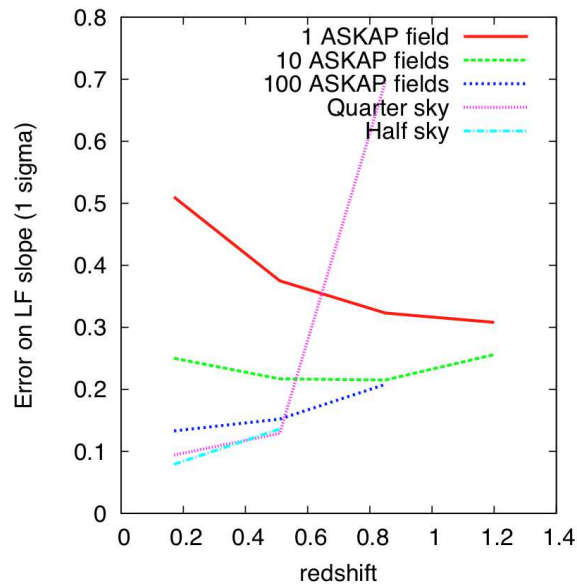} 
\caption{The uncertainty in the slope of the derived luminosity function for various sky areas. Note that the total survey time remains constant. Therefore, an all-sky survey will not be as deep as a survey with a single pointing. The uncertainty has been calculated at four redshift bins for each survey scenario. Note that the 100 ASKAP pointing, quarter and half sky surveys do not go deep enough at high redshifts for the slope of the luminosity function to be determined with any precision.} 
\label{slope}
\end{center} 
\end{figure}

Our goal is to determine the density evolution of the OH maser luminosity function so that we can investigate the merger-rate of the Universe out to redshift 1.4. However, it is likely that there will be changes in the luminosity function due to luminosity evolution of the OH maser population with redshift. This is to be expected because the dust content and star-formation rates of galaxies are known to increase with redshift, possibly peaking around redshift 1 to 2. Since, $L_{OH} \propto L_{FIR}$$^{2.3}$, it is natural to assume that the luminosity distribution of the OH maser galaxies will change also, which in turn will change the shape of the OH maser luminosity function. Therefore, it would be incorrect to assume that any change in the luminosity function is due purely to a change in the number of OH maser galaxies as a function of redshift. To investigate this, we have estimated the uncertainty in the slope of the luminosity function for various survey strategies using ASKAP. We have tested five different strategies which go from a single deep ASKAP pointing to a half-sky survey. Again, we have used a total time of 2160 hours to complete the survey in each case. Our results are shown in Figure \ref{slope}, where we present the uncertainty on the slope of the luminosity function against redshift for the five different area--depth combinations.

A single, deep ASKAP pointing is not particularly useful since it does not have a sufficient sky volume to pick up enough OH maser galaxies over all redshifts to suitably constrain the slope of the luminosity function. Conversely, the ASKAP 100 pointing, quarter and half-sky surveys do not go deep enough to probe sufficiently far down the luminosity function to constrain the slope at higher redshifts. An ASKAP 10 pointing survey appears to match well the required sensitivity and survey volume to uniformly constrain the slope of the luminosity function out to redshift 1.4. Note that the predicted uncertainty of the slope of the luminosity function for an ASKAP 10 pointing survey is similar to that currently found from low-redshift pointed surveys. The best solution is a shallow quarter-sky survey to constrain the slope at low redshifts and an ASKAP 10 pointing survey to constrain the slope at the highest redshifts.

\section{Conclusions}
\label{conc}

We have presented new simulations of the potential numbers of OH maser galaxies that can be detected with APERTIF and ASKAP out to redshift $\sim$1.4. Our results suggest that with 3 months of observing time, it will be possible to cleanly differentiate between evolutionary models for the OH maser population. Given that these new facilities will be carrying out large all-sky blind surveys for H\,{\sc i} in the same frequency range that we would expect to find high redshift OH, it would require only a small additional amount of effort to expand the search to look for OH galaxies.

The largest uncertainty in our analysis comes from the poorly constrained local OH maser luminosity function at high luminosities ($\geq 10^{3.8}~L_{\odot}$). There are a number of avenues available to try and constrain the high-luminosity end of the luminosity function in the short term, which would further constrain the predictions made here. The first is to carry out blind searches of small fields with current instruments, for example, with the Giant Metrewave Radio Telescope, as suggested by Darling \& Giovanelli \cite{darling02b}. Another option is to observe gravitationally lensed ULIRGs to detect OH masers at even higher redshifts ($z>2.5$). Here, the lensing magnification can be used to detect faint lines which would otherwise not be observable without the magnification of the lens. This technique was recently used to detect the most distant water maser from a high redshift galaxy \cite{impellizzeri08}.

\end{document}